\documentclass[prd,twocolumn,showpacs,preprintnumbers,floatfix,amsmath,amssymb,amsfonts]{revtex4}

\usepackage{graphicx}
\usepackage{amsmath}
\usepackage{dcolumn}
\usepackage{epsfig}
\usepackage{psfrag}

\RequirePackage{xspace}

\usepackage{relsize}
\def\babar{\mbox{\slshape B\kern-0.1em{\smaller A}\kern-0.1em
    B\kern-0.1em{\smaller A\kern-0.2em R}}}

\def\epem       {\ensuremath{e^+e^-}\xspace}

\def\ellell     {\ensuremath{\ell^+ \ell^-}\xspace}

\def\ccbar {\ensuremath{c\overline c}\xspace}

\def\piz   {\ensuremath{\pi^0}\xspace}

\def\ppz   {\ensuremath{\pi^0\pi^0}\xspace}

\def\pipi  {\ensuremath{\pi^+\pi^-}\xspace}

\def\Kbar  {\kern 0.2em\overline{\kern -0.2em K}{}\xspace}

\def\Kz    {\ensuremath{K^0}\xspace}
\def\Kzb   {\ensuremath{\Kbar^0}\xspace}
\def\KzKzb {\ensuremath{\Kz \kern -0.16em \Kzb}\xspace}
\def\Kp    {\ensuremath{K^+}\xspace}
\def\Km    {\ensuremath{K^-}\xspace}

\def\KpKm  {\ensuremath{\Kp \kern -0.16em \Km}\xspace}
\def\KS    {\ensuremath{K^0_{\scriptscriptstyle S}}\xspace} 
\def\KL    {\ensuremath{K^0_{\scriptscriptstyle L}}\xspace} 
\def\Kstarz  {\ensuremath{K^{*0}}\xspace}

\def\Kstar   {\ensuremath{K^*}\xspace}

\def\Kstarp  {\ensuremath{K^{*+}}\xspace}

\def\Kstarpm {\ensuremath{K^{*\pm}}\xspace}

\def\Kstarzmass  {\ensuremath{K^*\mskip-2mu(892)^{0}}\xspace}
\def\Kstarpmass  {\ensuremath{K^*\mskip-2mu(892)^+}\xspace}

\def\Dbar    {\kern 0.2em\overline{\kern -0.2em D}{}\xspace}

\def\Dz      {\ensuremath{D^0}\xspace}
\def\Dzb     {\ensuremath{\Dbar^0}\xspace}
\def\DzDzb   {\ensuremath{\Dz {\kern -0.16em \Dzb}}\xspace}
\def\Dp      {\ensuremath{D^+}\xspace}
\def\Dm      {\ensuremath{D^-}\xspace}

\def\DpDm    {\ensuremath{\Dp {\kern -0.16em \Dm}}\xspace}

\def\Dstarm  {\ensuremath{D^{*-}}\xspace}

\def\B       {\ensuremath{B}\xspace}
\def\Bbar    {\kern 0.18em\overline{\kern -0.18em B}{}\xspace}

\def\BB      {\ensuremath{B\Bbar}\xspace} 
\def\Bz      {\ensuremath{B^0}\xspace}
\def\Bzb     {\ensuremath{\Bbar^0}\xspace}
\def\BzBzb   {\ensuremath{\Bz {\kern -0.16em \Bzb}}\xspace}
\def\Bu      {\ensuremath{B^+}\xspace}
\def\Bub     {\ensuremath{B^-}\xspace}
\def\Bp      {\ensuremath{\Bu}\xspace}

\def\BpBm    {\ensuremath{\Bu {\kern -0.16em \Bub}}\xspace}

\def\jpsi     {\ensuremath{{J\mskip -3mu/\mskip -2mu\psi\mskip 2mu}}\xspace}
\def\psitwos  {\ensuremath{\psi{(2S)}}\xspace}

\def\etac     {\ensuremath{\eta_c}\xspace}

\def\chicone  {\ensuremath{\chi_{c1}}\xspace}

\mathchardef\Upsilon="7107
\def\Y#1S{\ensuremath{\Upsilon{(#1S)}}\xspace}
\def\FourS {\Y4S}
\newcommand{\gev}{\ensuremath{\mathrm{\,Ge\kern -0.1em V}}\xspace}
\newcommand{\mev}{\ensuremath{\mathrm{\,Me\kern -0.1em V}}\xspace}
\newcommand{\kev}{\ensuremath{\mathrm{\,ke\kern -0.1em V}}\xspace}
\newcommand{\ev}{\ensuremath{\mathrm{\,e\kern -0.1em V}}\xspace}
\newcommand{\gevc}{\ensuremath{{\mathrm{\,Ge\kern -0.1em V\!/}c}}\xspace}
\newcommand{\mevc}{\ensuremath{{\mathrm{\,Me\kern -0.1em V\!/}c}}\xspace}
\newcommand{\gevcc}{\ensuremath{{\mathrm{\,Ge\kern -0.1em V\!/}c^2}}\xspace}
\newcommand{\mevcc}{\ensuremath{{\mathrm{\,Me\kern -0.1em V\!/}c^2}}\xspace}
\def\mrad{\ensuremath{\rm \,mrad}\xspace} 

\def\invfb   {\ensuremath{\mbox{\,fb}^{-1}}\xspace}

\def\ps         {\ensuremath{{\rm \,ps}}\xspace}  

\def\to                 {\ensuremath{\rightarrow}\xspace}
\newcommand{\stat}{\ensuremath{\mathrm{(stat)}}\xspace}
\newcommand{\syst}{\ensuremath{\mathrm{(syst)}}\xspace}
\def\pep2{PEP-II}

\def\CP                {\ensuremath{C\!P}\xspace}
\def\stwob{\ensuremath{\sin\! 2 \beta   }\xspace}
\def\mistag{\ensuremath{w}\xspace}

\def\deltaz{\ensuremath{{\rm \Delta}z}\xspace}
\def\deltat{\ensuremath{{\rm \Delta}t}\xspace}
\def\deltamd{\ensuremath{{\rm \Delta}m_d}\xspace}
\newcommand{\jprlBase}       {Phys.\ Rev.\ Lett.\xspace}
\newcommand{\jprBase}        {Phys.\ Rev.\xspace}

\newcommand{\nimBaseA}       {Nucl.\ Instrum.\ Methods Phys.\ Res., Sect.\ A\xspace}
\newcommand{\npBase}         {Nucl.\ Phys.\xspace}
\newcommand{\zpBase}         {Z.\ Phys.\xspace}
\newcommand{\jpg}       [1]  {{J.\ Phys.\ {\bf G{\bf #1}}}}
\newcommand{\nima}      [1]  {\nimBaseA~{\bf #1}}
\newcommand{\np}        [1]  {\npBase\ {\bf #1}}
\newcommand{\jprl}      [1]  {\jprlBase\ {\bf #1}}
\newcommand{\jprd}      [1]  {\jprBase\ D~{\bf #1}}
\newcommand{\zpc}       [1]  {\zpBase\ C~{\bf #1}}

\newcommand{\e}      [1]   { {\ensuremath{ \times 10^{ {#1} } }}}

\def\etal     {{\it et al.}}

\newcommand\vud {\ensuremath{V_{\mathrm{ud}}}}

\newcommand\vub {\ensuremath{V_{\mathrm{ub}}}}
\newcommand\vcd {\ensuremath{V_{\mathrm{cd}}}}

\newcommand\vcb {\ensuremath{V_{\mathrm{cb}}}}
\newcommand\vtd {\ensuremath{V_{\mathrm{td}}}}

\newcommand\vtb {\ensuremath{V_{\mathrm{tb}}}}

\def\piz {\ensuremath{\pi^0}\xspace}
\def\jpsiks{\ensuremath{\jpsi\KS}\xspace}
\def\jpsikl{\ensuremath{\jpsi\KL}\xspace}
\def\jpsikz{\ensuremath{\jpsi\Kz}\xspace}
\def\psitwosks{\ensuremath{\psitwos\KS}\xspace}
\def\etacks{\ensuremath{\etac\KS}\xspace}
\def\chiconeks{\ensuremath{\chicone\KS}\xspace}
\def\jpsikstzero{\ensuremath{\jpsi\Kstarz}\xspace}
\def\jpsikstzeromass{\ensuremath{\jpsi\Kstarzmass}\xspace}
\def\JpsiKsCh{\ensuremath{\jpsiks(\pi^+\pi^-)}\xspace}
\def\JpsiKszz{\ensuremath{\jpsiks(\pi^0\pi^0)}\xspace}

\def\cpmodelist{\jpsiks, \jpsikl, \psitwosks, \etacks, \chiconeks, and \jpsikstzeromass}

\providecommand{\tbline}{\noalign{\vskip 0.04truecm\hrule\vskip0.04truecm}}
\newcommand\dbline{\noalign{\vskip 0.10truecm\hrule}\noalign{\vskip 2pt}  \noalign{\hrule\vskip 0.10truecm}}
\def\mes   {\ensuremath{m_\mathrm{ES}}\xspace}
\def\deltae{\ensuremath{\Delta E}\xspace}

\def\bb   {\ensuremath{B \overline{B}}}
\def\ifb   {\ensuremath{\mbox{\,fb}^{-1}}\xspace}

\def\Imlambda{\ensuremath{\mathop{\cal I\mkern -2.0mu\mit m}\lambda_f}}
\def\abslambda{\ensuremath{|\lambda_f|}\xspace}
\def\Btag{\ensuremath{B_\mathrm{tag}}\xspace}
\def\Brec{\ensuremath{B_\mathrm{rec}}\xspace}
\def\BCP{\ensuremath{B_{\CP}}\xspace}
\def\Bflav{\ensuremath{B_\mathrm{flav}}\xspace}

\def\Gammad{\ensuremath{\Gamma_d}\xspace}
\def\deltaGammad{\ensuremath{\Delta \Gammad}\xspace}
\def\Abar    {\kern 0.20em\overline{\kern -0.20em A}{}\xspace}
\def\Ab      {\ensuremath{\Abar}\xspace}

\def\effectiveeta{0.504 \pm 0.033}
\def\fitr{0.233}\def\statr{0.010}\def\systr{0.005}
\def\measurerperp{\fitr \pm \statr\, \stat \pm \systr\, \syst}

\def\lumi {\ensuremath{425.7\, \ifb}}
\def\nbb {\ensuremath{(465\pm 5)\e{6}\, \bb\ \rm pairs}}
\def\nbbx {\ensuremath{(465\pm 5)\e{6}}}
\def\extralumi{211.7}

\def\fittedS{\ensuremath{0.687 \pm 0.028}}
\def\fittedC{\ensuremath{0.024 \pm 0.020}}
\def\rhoSC{\ensuremath{0.1}}
\def\systS{\ensuremath{0.012}}
\def\systC{\ensuremath{0.016}}
\def\fittedstwob{\ensuremath{0.687 \pm 0.028}}
\def\fittedmodl{\ensuremath{0.977 \pm 0.020}}
\def\fittedSks{\ensuremath{0.662 \pm 0.039}}
\def\fittedSkszz{\ensuremath{0.625 \pm 0.091}}
\def\fittedSpsitwoS{\ensuremath{0.897 \pm 0.100}}
\def\fittedSchic{\ensuremath{0.614 \pm 0.160}}
\def\fittedSetac{\ensuremath{0.925 \pm 0.160}}
\def\fittedSkl{\ensuremath{0.694 \pm 0.061}}
\def\fittedSkst{\ensuremath{0.601 \pm 0.239}}
\def\fittedCks{\ensuremath{0.017 \pm 0.028}}
\def\fittedCkszz{\ensuremath{0.091 \pm 0.063}}
\def\fittedCpsitwoS{\ensuremath{0.089 \pm 0.076}}
\def\fittedCchic{\ensuremath{0.129 \pm 0.109}}
\def\fittedCetac{\ensuremath{0.080 \pm 0.124}}
\def\fittedCkl{\ensuremath{-0.033 \pm 0.050}}
\def\fittedCkst{\ensuremath{0.025 \pm 0.083}}
\def\fittedSjpsikz{\ensuremath{0.666 \pm 0.031}}
\def\fittedCjpsikz{\ensuremath{0.016 \pm 0.023}}
\def\fittedSjpsiks{\ensuremath{0.657 \pm 0.036}}
\def\fittedCjpsiks{\ensuremath{0.026 \pm 0.025}}
\def\measuredc{\ensuremath{\fittedC\, \stat \pm \systC\, \syst}}
\def\measureds{\ensuremath{\fittedS\, \stat \pm \systS\, \syst}}

\begin{document}
\begin{flushleft}
SLAC-PUB-13544 \\
\babar-PUB-08/055
\end{flushleft}

\title{\boldmath Measurement of Time-Dependent \CP Asymmetry in $\Bz\to\ccbar K^{(*)0}$ Decays.}

%
\author{B.~Aubert}
\author{Y.~Karyotakis}
\author{J.~P.~Lees}
\author{V.~Poireau}
\author{E.~Prencipe}
\author{X.~Prudent}
\author{V.~Tisserand}
\affiliation{Laboratoire d'Annecy-le-Vieux de Physique des Particules (LAPP), Université de Savoie, CNRS/IN2P3, F-74941 Annecy-Le-Vieux, France }
\author{J.~Garra~Tico}
\author{E.~Grauges}
\affiliation{Universitat de Barcelona, Facultat de Fisica, Departament ECM, E-08028 Barcelona, Spain }
\author{L.~Lopez$^{ab}$ }
\author{A.~Palano$^{ab}$ }
\author{M.~Pappagallo$^{ab}$ }
\affiliation{INFN Sezione di Bari$^{a}$; Dipartmento di Fisica, Universit\`a di Bari$^{b}$, I-70126 Bari, Italy }
% ...

\collaboration{The \babar\ Collaboration}
\noaffiliation

\begin{abstract}
We present updated measurements of time-dependent \CP\ asymmetries in fully reconstructed
neutral \B\ decays containing a charmonium meson.  The measurements reported here use
a data sample of \nbbx\ $\Upsilon(4S) \to \BB$ decays collected with the \babar\ detector
at the \pep2\ asymmetric energy \epem\ storage rings
operating at the SLAC National Accelerator Laboratory.
The time-dependent \CP\ asymmetry parameters measured from
\cpmodelist
decays are: 
$C_f = \measuredc$ and  
$-\eta_f S_f = \measureds$ .
\end{abstract}
\pacs{13.25.Hw, 12.15.Hh, 11.30.Er}
                                                                                
\maketitle

\section{INTRODUCTION}
\label{sec:intro}

The Standard Model (SM) of electroweak interactions describes
\CP violation as a consequence of an
irreducible phase in the three-family Cabibbo-Kobayashi-Maskawa (CKM)
quark-mixing matrix~\cite{ref:CKM}.
In the CKM framework, tree-diagram 
processes dominate neutral \B\ decays to \CP\ eigenstates containing a
charmonium and a $K^{(*)0}$ meson. These provide a direct measurement of
\stwob~\cite{BCP}, where the angle $\beta$ is defined in terms of the
CKM matrix 
elements $V_{\mathrm{ij}}$ for quarks i, j  
as $\arg [-(\vcd^{}\vcb^*) / (\vtd^{}\vtb^*)]$.

We identify (tag) the initial flavor of the reconstructed \B
candidate, \Brec, using information from the other \B meson, \Btag, in the
event.
The decay rate $g_+$ $(g_-)$ for a neutral \B meson to a \CP
eigenstate $f$ accompanied by a \Bz (\Bzb) tag, before taking into account detector
resolution effect, can be expressed as
\begin{eqnarray}
\label{eq:timedist}\nonumber
g_\pm(\deltat) &&= \frac{e^{{- | \deltat |}/\tau_{\Bz} }}{4\tau_{\Bz} }
\left\{ (1\mp\Delta\mistag)  \pm  (1-2\mistag)
\times \right. \\
&&\left.\left[ S_f\sin(\deltamd\deltat) -
  C_f\cos(\deltamd\deltat)  \right] \right\}\:\:\:
\end{eqnarray}
where
\begin{eqnarray}
S_f &=& \frac{2\Imlambda}{1+\abslambda^2},\nonumber\\
C_f &=& \frac{1 - \abslambda^2 } {1 + \abslambda^2},\nonumber
\end{eqnarray}
$\deltat \equiv t_\mathrm{rec} - t_\mathrm{tag}$ is the difference
between the proper decay times of \Brec and \Btag,
$\tau_{\Bz}$ is the neutral \B lifetime, and \deltamd is the mass difference
between the \B meson mass eigenstates
determined from $\Bz$-$\Bzb$ oscillations~\cite{Yao:2006px}.
Here, $\lambda_f=(q/p)(\Ab_f/A_f)$~\cite{ref:lambda},
where $q$ and $p$ are complex constants that relate the \B-meson flavor
eigenstates to the mass eigenstates, and $\Ab_f/A_f$ is the ratio of the
$\Bzb$ and $\Bz$ decay amplitudes to the final state $f$.
We assume that
the corresponding decay-width difference \deltaGammad is zero.
The average mistag probability \mistag describes the effect of incorrect
tags and $\Delta\mistag$ is the difference between the mistag probabilities
for \Bz and \Bzb\ mesons.
The sine term in Eq.~\ref{eq:timedist} results from the interference
between direct decay and decay after $\Bz$-$\Bzb$ oscillation. A
non-zero cosine term arises from the interference between decay amplitudes
with different weak and strong phases (direct \CP violation $|\Ab_f/A_f|\neq 1$) 
or from \CP
violation in $\Bz$-$\Bzb$ mixing ($|q/p|\neq 1$).
In the SM, \CP violation in mixing and
direct \CP violation are both
negligible in $b\to\ccbar s$ decays~\cite{ref:lambda}. 
Under these assumptions, $\lambda_f=\eta_f e^{-2i\beta}$, 
where $\eta_f=+1$ ($-1$) is the \CP eigenvalue for a \CP-even (odd)
final state, 
implying $C_f=0$. Thus, the time-dependent \CP-violating asymmetry is
\begin{eqnarray}
A_{\CP}(\deltat) &\equiv& \frac{g_+(\deltat) - g_-(\deltat)}{g_+(\deltat) +
g_-(\deltat)} \\ \nonumber
&=& (1-2\mistag)S_f \sin{ (\deltamd \, \deltat )},
\label{eq:asymmetry}
\end{eqnarray}
and $S_f=-\eta_f\stwob$.  If the assumption that $C_f=0$ is relaxed, then
$S_f = -\eta_f \sqrt{1-C_f^2}\stwob$.

In a previous publication~\cite{Aubert:2007hm},  
we reported time-dependent \CP asymmetries
in terms of the parameters \stwob and \abslambda.  In this paper we
report results in terms of $S_f$ and $C_f$ to be consistent with 
other time-dependent \CP asymmetry measurements.
We reconstruct \Bz decays to the final states
\jpsiks, \jpsikl, \psitwosks, \chiconeks, \etacks, and
\jpsikstzeromass with $\Kstarzmass\to \KS \pi^0$~\cite{ref:chargeconj}.  
The $\jpsikl$ final state is
\CP-even  and the $\jpsikstzero$ final state is an admixture of \CP-even
and \CP-odd amplitudes.
The remaining final states are \CP-odd.
The \CP-even and odd amplitudes in $\Bz\to\jpsikstzero$ decays can be separated in 
an angular analysis~\cite{Aubert:2007hz}.
In this analysis, we average over the angular information resulting in a dilution
of the measured \CP asymmetry by a factor $1-2R_{\perp}$, where
$R_{\perp}$ is the fraction of the $L$=1 contribution.
In Ref.~\cite{Aubert:2007hz} we have measured $R_{\perp} = \measurerperp$,
which gives an effective $\eta_f = \effectiveeta$ after acceptance corrections
for $f=\jpsikstzero$.  In addition to measuring 
a combined $S_f$ and $C_f$ for the
\CP modes described above, we measure $S_f$ and $C_f$ for each final
state $f$ individually.
We split the $\jpsiks$ mode into samples with either $\KS\to
\pi^+\pi^-$ or $\pi^0\pi^0$. We also combined $\jpsikz$ channel with
$\Kz$ either a $\KS$ or $\KL$.
Compared to our previous publication~\cite{Aubert:2007hm}, the current
analysis contains 
$82\times10^6$ additional $\BB$ decays and improved track 
reconstruction algorithms have been applied to the entire data set.

\section{\boldmath THE DATA SET AND \babar\ DETECTOR\label{sec:dataset}}
The results presented in this paper are based on data collected 
with the \babar\ detector at the \pep2\ asymmetric energy \epem\ storage 
rings~\cite{ref:pepcdr}
 operating at the SLAC National Accelerator Laboratory. At \pep2, 
9.0 \gev\ electrons and 3.1 \gev\ positrons collide at a center-of-mass
energy of 10.58 \gev, which corresponds to the mass of the \FourS\ resonance.  
The asymmetric energies result in a boost from the 
center-of-mass (CM) frame to the laboratory of $\beta\gamma\approx 0.56$.
The data set analyzed has an integrated luminosity of \lumi\
corresponding to \nbb\ recorded at the \FourS\ resonance. 

The \babar\ detector is described in detail elsewhere~\cite{Aubert:2001tu}.
Surrounding the interaction point is a five-layer, double-sided
silicon vertex tracker (SVT), which measures the impact parameters of 
charged particle tracks in both the plane transverse to, and along, 
the beam direction. A 40-layer drift chamber surrounds the SVT 
and provides measurements of the momenta for charged 
particles. 
Charged hadron identification is achieved 
through measurements of particle energy-loss in the tracking system 
and the Cherenkov angle obtained from a detector of internally 
reflected Cherenkov light. A CsI(Tl) electromagnetic calorimeter 
(EMC) provides photon detection, electron identification, and 
$\piz$ reconstruction. 
The aforementioned components are enclosed by a solenoid magnet, 
which provides
a 1.5 T magnetic field.
Finally, the flux return of 
the magnet (IFR) is instrumented in order to allow discrimination of
muons from pions.  
For the most recent $\extralumi\invfb$ of data, a portion of the 
resistive plate chambers in the IFR 
has been replaced by limited streamer tubes ~\cite{Menges:2006xk}.

We use a right-handed coordinate system with the $z$ axis along the
electron beam direction and the $y$ axis upward. Unless otherwise
stated, kinematic quantities are calculated in the laboratory rest
frame. 
We use Monte Carlo (MC) simulated events generated with the
\babar\ simulation based on GEANT4~\cite{Agostinelli:2002hh} for detector
responses and EvtGen~\cite{Lange:2001uf} for event kinematics
to determine signal and background
characteristics, optimize selection criteria, and evaluate
efficiencies.

\section{\boldmath RECONSTRUCTION OF \B\ CANDIDATES\label{sec:reconstruction}}

We select two samples of events in order to measure the time-dependent \CP\ asymmetry
parameters $S_f$ and $C_f$: a sample of signal events used in the extraction
of the \CP\ parameters (\BCP) and a sample of fully reconstructed \B\ meson decays
to flavor eigenstates (\Bflav).
The \BCP sample consists of 
\Bz decays to \cpmodelist, where \Kstarz decays to $\KS\piz$.
The \Bflav\ sample consists of \Bz\ decays to
$D^{(*)-}(\pi^+,\,\rho^+,\,a_1^+)$ final
states.  We use the \Bflav sample to determine the
dilution (mistag probability) and the resolution function,
discussed in Section~\ref{sec::fit}.
We assume that the interference between the \CP side and the tag side
reconstruction is negligible and therefore that the dilution 
and resolution parameters are the same for the \Bflav and \BCP samples. 
We also select a sample of fully reconstructed
charged \B\ meson decays to $\jpsi K^+$, $\psitwos K^+$,
$\chicone K^+$, $\etac K^+$, and $\jpsi \Kstarpmass$, where $\Kstarp$
decays to $K^+\piz$ or $\KS\pi^+$,  to use as a
control sample.

The event selection is unchanged
from that described in Ref~\cite{Aubert:2007hm}.
$\jpsi$ and $\psitwos$ mesons are reconstructed via their decays to
\epem or $\mu^+\mu^-$ final states. 
At least one of the
leptons is required to pass a likelihood particle identification
algorithm based on the information provided by the EMC, the IFR and 
from ionization
energy loss measured in the tracking system. We require the invariant mass 
of the muon pair $m(\mu^+\mu^-)$ to be in the mass range 
3.06--3.14\gevcc for \jpsi or 3.636--3.736\gevcc for
\psitwos candidates.
For $\jpsi\to \epem$ and $\psitwos\to \epem$ decays, where the
electron may 
have radiated bremsstrahlung photons, part of the missing energy 
is recovered by identifying neutral clusters with more than $30\,\mev$
lying within $35\,\mrad$ in polar angle and $50\,\mrad$ in azimuth of the 
electron direction projected onto EMC. 
The invariant mass of \epem pairs  is required to be within
2.95--3.14\gevcc for \jpsi candidates,
or 3.436--3.736\gevcc for \psitwos candidates.

We also construct \psitwos\ mesons in the $\jpsi\pi^+\pi^-$ final
state, where the \jpsi candidate is combined with a pair of
oppositely-charged tracks assumed as pions with no particle
identification applied, and the pion pair-invariant mass between
400\mevcc and 600\mevcc. Candidates with $3.671\,\gevcc <
m(\jpsi\pi^+\pi^-) < 3.701\,\gevcc$ are retained.

The \chicone\ candidates are reconstructed in the $\jpsi\gamma$
final state.
The photon candidates  are required to have an energy greater than 
100\mev but less than 2\gev, and, when combined with other
photons, not to form a \piz\ candidate with invariant mass $120\,\mevcc
< m(\gamma\gamma) < 150\,\mevcc$.
The invariant mass of the
$\chicone$ candidate is  required to be between 3.477\gevcc and 
3.577\gevcc. Mass constraints are applied in the fits 
to improve the determinations of the energies and momenta of the 
\jpsi, \psitwos, and \chicone candidates.

We reconstruct the $\Bz\to\etac\KS$ mode using 
the $\etac\to\KS K^+\pi^-$ decay. We exploit the fact that the
$\etac$ decays predominantly through a $K\pi$ resonance at around 
$1.43\,\gevcc$ and a $\KS K$ resonance close to the threshold. We
require that $m(\KS\pi^-)$ or $m(K^+\pi^-)$ is within mass range of
$1.26\,\gevcc$ and $1.63\,\gevcc$,
or $1.0\,\gevcc < m(K^+\KS) < 1.4\,\gevcc$.

The decay channels $K^+\pi^-$, $K^+\pi^-\piz$,
$K^+\pi^+\pi^-\pi^-$, and $\KS\pi^+\pi^-$ are used to reconstruct $\Dzb$,
while $D^-$ candidates are selected in the $K^+\pi^-\pi^-$ and 
$\KS\pi^-$ modes. 
We require that the $\Dzb$ and $D^-$ candidate invariant mass is
within $\pm3\sigma$ of their respective nominal mass,
where $\sigma$ is the uncertainty calculated for each candidate.
A mass-constrained fit is then applied to 
the $\Dzb$ and $D^-$ candidates satisfying these requirements.
We form $\Dstarm$ candidates in the decay $\Dstarm\to\Dzb\pi^-$ by
combining a $\Dzb$ with a pion that has momentum greater than $70\,\mevc$.
The $\Dstarm$ candidates are required to have $m(\Dzb\pi^-)$ within
$\pm1.1\,\mevcc$ of the nominal $\Dstarm$ mass for the
$\Dzb\to K^+\pi^-\piz$ mode and $\pm0.8\,\mevcc$ for all other
modes. 

For the \jpsiks\ decay, we use both $\KS \to\pi^+\pi^-$
and $\KS\to\piz\piz$ decays; for other \B decay modes we only use
$\KS\to\pi^+\pi^-$. 
Candidates in the $\KS\to\pi^+\pi^-$ mode are selected by requiring an
invariant $\pi^+\pi^-$ mass, computed at the vertex of the two
oppositely-charged tracks, between $472.67\,\mevcc$ and $522.67\,\mevcc$.
We further apply a mass constraint fit to the $\KS$ candidates
before combining them with charmonium candidates to form 
$\Bz$ candidates. Neutral pion candidates, in the mass range 
100--155\mevcc, are formed from two $\gamma$ candidates from 
the EMC. Pairs of \piz\ are combined to construct $\KS\to\piz\piz$ candidates. 
The minimum energy is required to be 30\mev for $\gamma$, 200\mev 
for $\piz$, and 800\mev for $\KS$ candidates.
To select $\KS$ candidates, the $\piz\piz$ invariant mass 
is restricted to the region between 470\mevcc and 550\mevcc.

Candidates for \KL\ are identified in the EMC and IFR detectors as 
reconstructed clusters that cannot be associated with any charged track
in the event.
As the energy of $\KL$ cannot be measured well, the laboratory momentum of
the \KL\ is determined by its flight direction and the constraint that the
invariant mass of the $\jpsi\KL$ system has the known \Bz\ mass.  
For events with multiple $\jpsi\KL$ candidates, a hierarchy is imposed where 
the highest energy EMC cluster for multiple EMC combinations, or the
IFR cluster with the largest number of layers
for multiple IFR combinations, is selected. In case both EMC and IFR
combinations are found, the EMC combination is chosen because of its better
angular resolution.

 We reconstruct $K^{*0}$ candidates in the 
$\KS\piz$ mode, while $\Kstarp$ candidates are reconstructed in the
$K^+\piz$ and $\KS\pi^+$ modes. The invariant mass of the two daughters is
required to be within $\pm100\,\mevcc$ of the nominal $\Kstar$ mass.

The $\rho^+$ candidates are reconstructed in the $\pi^+\piz$ final 
state, where the $\pi^+\piz$ mass is required to lie within $\pm150\,\mevcc$ 
of the nominal $\rho^+$ mass. 
Candidates in the decay mode $a^+_1\to\pi^+\pi^-\pi^+$ are reconstructed
by combining three charged tracks with pion mass assumption, and
restricting the three-pion invariant mass to lie between 1.0 and 1.6\gevcc. 

Events that pass the selection requirements are refined using
kinematic variables.  For the \jpsikl\ mode, the difference \deltae\ 
between the candidate CM energy and the beam energy
in the CM frame, $E^*_\mathrm{beam}$, is required to satisfy $|\deltae| < 80\mev$.
For all other categories of events,
we require $|\deltae| <20\,\mev$ and the beam-energy substituted mass 
$\mes=\sqrt{(E^*_\mathrm{beam})^2-(p^*_B)^2}$
to be greater than $5.2\gevcc$, where $p^*_B$ is the \B\ momentum in the CM frame. 
When multiple $B$ candidates (with $\mes>5.2\,\gevcc$) are found in the
same event, the candidate with the smallest value of $|\deltae|$ is 
selected.

We calculate the proper  time difference 
\deltat between the two \B decays from the
measured separation \deltaz between the decay vertices of \Brec and \Btag
along the collision ($z$) axis \cite{Aubert:2002rg}.
The $z$ position of the \Brec vertex is determined from the charged
daughter tracks. The \Btag decay vertex is determined by fitting tracks not
belonging to the \Brec candidate to a common vertex,
including constraints from the beam spot location and the
\Brec momentum~\cite{Aubert:2002rg}.
Events are accepted if the calculated $\deltat$ uncertainty is less than
$2.5\ps$ and $|\deltat|$ is less than $20\ps$. The fraction of signal
MC events satisfying such a requirement is $95\,\%$.

\section{\boldmath \B\ MESON FLAVOR TAGGING\label{sec:tagging}}

A key ingredient in the measurement of time-dependent \CP\ asymmetries
is the determination of whether the \Brec\ was a \Bz\ or a
\Bzb\ at the time of $\Delta t =0$. This `flavor tagging'
is achieved with the analysis of the decay products of the recoiling \B\ meson \Btag.
The overwhelming majority of \B\ mesons decay to a final state that is flavor-specific, i.e.,
only accessible from either a \Bz\ or a \Bzb. The purpose of the flavor-tagging
algorithm is to determine the flavor of \Btag\ with the highest efficiency $\epsilon_{\rm tag}$
and lowest probability \mistag\ of assigning the wrong flavor.  It is not necessary
to fully reconstruct \Btag\ in order to determine its flavor.

The figure of merit for the performance of the tagging algorithm is the effective 
tagging efficiency
\begin{equation}
Q = \epsilon_{\rm tag} (1-2\mistag)^2,
\end{equation}
which is related to the statistical uncertainty $\sigma_S$ and $\sigma_C$ in the coefficients $S_f$ and $C_f$ through
\begin{equation}
\sigma_{S,C} \propto \frac{1}{\sqrt{Q}}.
\end{equation}

The tagging algorithm we employ~\cite{Aubert:2002rg,Aubert:2007hm}  
analyzes tracks 
on the tag side to assign a flavor and associated probability to \Btag. The flavor of
\Btag\ is determined from a combination of nine different tag signatures, such as 
isolated primary leptons, kaons and pions from \B\ decays to final states containing $D^*$
mesons, and high momentum charged particles from \B\ decays.
The properties of those signatures are used as inputs to a single neural network that is trained to
assign the correct flavor to \Btag. The output of this neural network then  
is divided into seven mutually-exclusive categories.
These are (in order of decreasing signal purity)  
{\it Lepton, Kaon I, Kaon II, KaonPion,
Pion, Other} and {\it Notag}. 
The events with the neural network output $|NN|>0.8$ are defined as
{\it Lepton} category, if they are also accompanied by an isolated
primary lepton; otherwise they are categorized as {\it Kaon I} tag. For
the other five tag categories ({\it Kaon II, KaonPion, Pion, Other} and
{\it Notag}) the outputs of the neutral network are required to satisfy:
$0.6<|NN|<0.8$, $0.4<|NN|<0.6$, $0.2<|NN|<0.4$, $0.1<|NN|<0.2$, and
$|NN|<0.1$, respectively.

The performance of this algorithm is evaluated using the \Bflav\ sample. The
final state of the \Bflav\ sample can be classified as mixed or unmixed depending on
whether the reconstructed flavor-eigenstate $\Bflav$ has the same or opposite
flavor as the tagging $\B$. After taking the mistag probability into account,
the decay rate $g_{\pm,\Bz}$  ($g_{\pm,\Bzb}$) 
for a neutral \B\ meson to decay to a 
flavor eigenstate accompanied by a \Bz\ (\Bzb) tag can be expressed as
\begin{eqnarray}
\label{eq:timedist_bflav} 
g_{\pm,\Bz}(\deltat) &\propto& [(1-\Delta\mistag_i)\pm(1-2\mistag_i)\cos(\deltamd\deltat) ] 
\nonumber \\\\ \nonumber
g_{\pm,\Bzb}(\deltat) &\propto& [(1+\Delta\mistag_i)\pm(1-2\mistag_i)\cos(\deltamd\deltat) ], 
\end{eqnarray}
where the $\pm$ sign in the index refers to mixed $(-)$ and unmixed
$(+)$ events; the index $i$ denotes the $i$th tagging category.
The performance of the tagging algorithm is summarized in Table~\ref{tab:mistag}.
The events in the {\it Notag} category contain no flavor information, so
carry no weight in the 
time-dependent analysis. They are excluded from further analysis. 
The total effective tagging efficiency
is measured to be $(31.2\pm 0.3)\%$.
\begin{table}[!ht]
\begin{center}
\caption{Efficiencies $\epsilon_i$, average mistag fractions $\mistag_i$, 
mistag fraction differences between \Bz\ and \Bzb\ tagged events $\Delta\mistag_i$, 
and effective tagging
efficiency $Q_i$ extracted for each tagging
category $i$ from the $B_{\rm flav}$ sample.\label{tab:mistag}}
\vspace{0.5cm}
\begin{tabular}{lcccc}\hline\hline
Category     &
{$\epsilon_i$   (\%)} &
{$\mistag_i$       (\%)} &
{$\Delta\mistag_i$ (\%)} &
{$\ \ \ Q_i$             (\%)} \\ \hline

{\it Lepton}  & $\phantom{0}8.96  \pm 0.07$ &  $\phantom{0}2.8\pm 0.3$ & $\phantom{-}0.3\pm 0.5$    & $\phantom{0}7.98\pm 0.11$ \\
{\it Kaon I}   & $10.82 \pm 0.07$ &  $\phantom{0}5.3\pm 0.3$ & $-0.1\pm 0.6$    & $\phantom{0}8.65\pm 0.14$ \\
{\it Kaon II}   & $17.19 \pm 0.09$ &  $14.5\pm 0.3$ & $\phantom{-}0.4\pm 0.6$  & $\phantom{0}8.68\pm 0.17$  \\
{\it KaonPion} & $13.67 \pm 0.08$ &  $23.3\pm 0.4$ & $-0.7\pm 0.7$ & $\phantom{0}3.91\pm 0.12$  \\
{\it Pion}    & $14.18 \pm 0.08$ &  $32.5\pm 0.4$ & $\phantom{0}5.1\pm 0.7$   & $\phantom{0}1.73\pm 0.09$  \\
{\it Other}  & $\phantom{0}9.54  \pm 0.07$ &  $41.5\pm 0.5$ & $\phantom{-}3.8\pm 0.8$  & $\phantom{0}0.27\pm 0.04$  \\\hline
All   & $74.37 \pm 0.10$ &                &               & $31.2\pm 0.3$ \\\hline\hline
\end{tabular}
\end{center}
\end{table}

\section{\boldmath LIKELIHOOD FIT METHOD\label{sec:maximum}}
\label{sec::fit}
We determine the composition of our final sample by performing simultaneous
fits to the \mes\ distributions for the full \BCP\ and \Bflav\ samples, except for the
\jpsikl\ sample for which we extract the \KL\ momentum by using the 
\Bz\ mass constraint and fit the \deltae\ distribution.
We then perform a simultaneous maximum likelihood
fit to the \deltat distribution of the tagged \BCP and \Bflav samples
to measure $S_f$ and $C_f$.

\begin{figure*}[!htb]
\begin{center}
\includegraphics[width=0.85\textwidth]{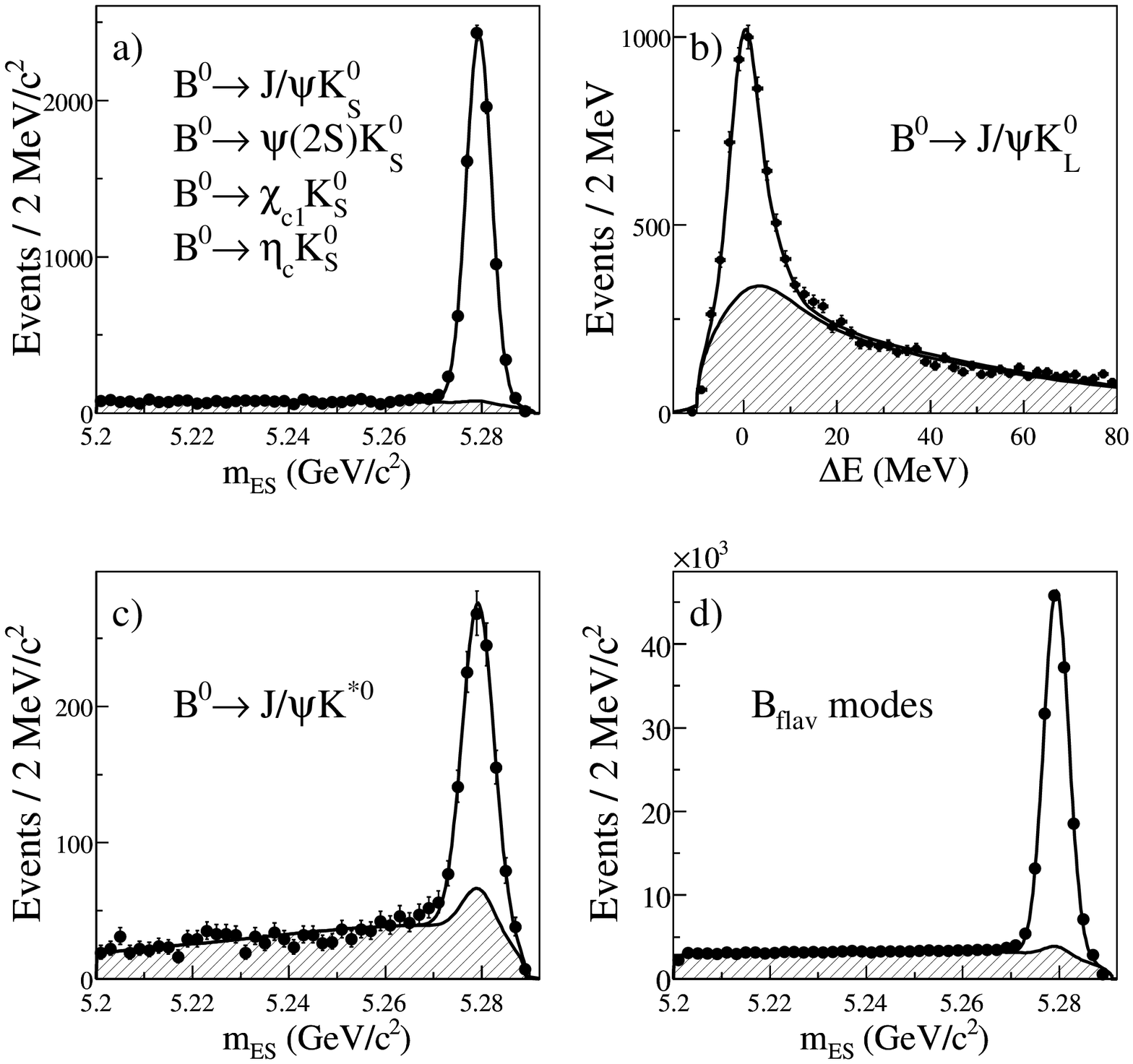}
\caption{
Distributions for \BCP and \Bflav candidates satisfying the tagging and
vertexing requirements:
a) \mes\ for the final states $\jpsi\KS $, $\psitwos\KS$, $\chicone\KS$,
   and $\etac\KS$,
b) \deltae for the final state $\jpsi\KL$,
c) \mes\ for $\jpsi\Kstarz(\Kstarz\to \KS\piz)$, and
d) \mes\ for the \Bflav sample. In each plot, the shaded region is the
   estimated background contribution.
}
\label{fig:bcpsample}
\end{center}
\end{figure*}

We define a signal region of $5.27 < \mes < 5.29 \gevcc$ 
($|\deltae|< 10\mev$ for $\jpsi\KL$), which contains 15481 candidate events 
of \BCP\ sample that satisfy the tagging
and vertexing requirements (see Table~\ref{tab:result}).
The signal \mes\ distribution for the full \BCP\ and \Bflav\ samples, except for the
\jpsikl\ sample is described by a Gaussian function.
The background \mes\ distribution is modeled by an ARGUS threshold 
function~\cite{Albrecht:1990am}, where 
a shape parameter is allowed to vary in the fit.
For the decay modes of $\jpsi\KS$, $\psi2s\KS$, $\chi_{c1}\KS$, $\jpsi K^{*0}$ and
$\Bflav$,  we use simulated events to
estimate the fractions of background events that peak in the \mes\ signal region 
($\mes>5.27\,\gevcc$) due to cross-feed from other decay modes.
We describe this component with a Gaussian function having the same mean
and width as the signal and refer to it as peaking background because
if neglected, it would lead to an overestimate of the signal yield.
The peaking background is less than 1\,\% in the decay of $\Bz\to\jpsi\KS$,
and at the level of a few percent in most other decay modes. 
The only exception is the decay of $\Bz\to\jpsi K^{*0}$, where the 
peaking background level is about $13\,\%$. MC simulations show that
it consists  of $44\,\%$ of $\Bp$ decays,
$32\,\%$ of $\Bz\to \chi_{c} \KS$ decays and $24\,\%$ of other $\Bz$ decays.
For the $\etac\KS$ mode, the cross-feed fraction is determined
from a fit to the $m_{KK\pi}$ and \mes\ distributions in data.
For the $\jpsi\KL$ decay mode, the signal \deltae\ distribution 
is determined from MC simulated events. The sample composition, effective $\eta_f$,
and \deltae\ distribution of the individual background sources are
determined either from simulation (for $B\to\jpsi X$) or from the
$m_{\ellell}$ sidebands in data (for non-\jpsi background).
Figure~\ref{fig:bcpsample} shows the distributions of
\mes\ obtained for the \BCP and \Bflav events, and \deltae\ obtained for 
the \jpsikl\ events.

The \deltat distributions of the \BCP sample are modeled by
Eq.~\ref{eq:timedist} and those of the \Bflav sample by
Eq.~\ref{eq:timedist_bflav}.
The \deltat distributions for the signal are convolved with a
resolution function common to both the \Bflav and \BCP samples,
modeled by the sum of three Gaussian functions~\cite{Aubert:2002rg}, 
called the core, tail and outlier components, which can be represented 
as a function of the reconstruction uncertainty 
$\delta t = \Delta t - \Delta t_\mathrm{true}$ 
as follows:
\begin{eqnarray}
\mathcal{R}(\delta t;\sigma_{\Delta t}) 
& = & f_\mathrm{core} h_G(\delta t;\delta_\mathrm{core}\sigma_{\Delta t},S_\mathrm{core}\sigma_{\Delta t}) \nonumber \\
& + & f_\mathrm{tail} h_G(\delta t;\delta_\mathrm{tail}\sigma_{\Delta t},S_\mathrm{tail}\sigma_{\Delta t}) \nonumber \\
& + & f_\mathrm{out} h_G(\delta t;\delta_\mathrm{out},S_\mathrm{out}), 
\end{eqnarray}
where
\begin{equation}
h_G(\delta t;\delta,\sigma) 
=  \frac{1}{\sqrt{2\pi}\sigma} 
\exp\left(-\frac{(\delta t - \delta)^2}{2\sigma^2}\right),
\end{equation}
and
\begin{equation}
f_\mathrm{core} + f_\mathrm{tail} + f_\mathrm{out} = 1.
\end{equation}
The widths ($\sigma$) of the core and tail components include two independent 
scale factors, $S_\mathrm{core}$ and $S_\mathrm{tail}$, to accommodate an overall
underestimate or overestimate of the $\Delta t$ measurement error $\sigma_{\Delta t}$
for all events. The parameter $S_\mathrm{core}$ is free in the fit and
its value is close to unity.
The value of $S_\mathrm{tail}$ is derived from MC studies and fixed to be 3.
Studies show that the measurement of $C_f$ and $S_f$ is not sensitive to
the choice of  the $S_\mathrm{tail}$ value. 
We later vary the $S_\mathrm{tail}$ value within
a large region and assign the shift of the measured $C_f$ and $S_f$ values as the
corresponding systematic uncertainties.  
We account for residual charm decay products included in the \Btag\ candidate
vertex by allowing the core and tail Gaussian functions to have non-zero
mean values (bias, $\delta_\mathrm{core}\neq 0$ and $\delta_\mathrm{tail}\neq 0$). 
The bias ($\delta_\mathrm{core}$) and width ($S_\mathrm{core}$) of 
the core component are allowed to
differ for the lepton-tagged and nonlepton-tagged events. 
We use common parameters for the tail component.  
In order to account for the strong correlations with other resolution
parameters, 
the outlier component bias ($\delta_\mathrm{out}$) and 
width ($S_\mathrm{out}$) are fixed to 0 ps and 8 ps, respectively. 

The \deltat spectrum of the combinatorial background is described 
by an empirical distribution, consisting of components with zero and non-zero 
lifetimes ($\tau_{\mathrm{bg}}$) 
that are convolved with a resolution function~\cite{Aubert:2002rg} 
distinct from that used for the signal. Here, we use a double-Gaussian 
distribution, which has components similar to the core and outlier 
distributions described above. In this case, the resolution function
 is common to all tagging categories. 
The peaking background is assigned
the same \deltat distribution as the signal but with $S_f=C_f=0$,
and uses the same \deltat resolution function as the signal.  The non-zero lifetime
component of the combinatorial background contains both mixed
and un-mixed events. Therefore we allow the value of $\deltamd$
for this component ($\Delta m_{d,\mathrm{bg}}$)
to vary in the fit.  

In addition to $S_f$ and $C_f$, there are 69 free parameters in the
fit. For the signal, these are 
\begin{itemize}
  \item 7 parameters for the \deltat resolution: $\delta_\mathrm{core}$ and 
  $S_\mathrm{core}$ for the lepton-tagged and nonlepton-tagged events, 
  $f_\mathrm{core}$, $f_\mathrm{tail}$, and $\delta_\mathrm{tail}$.

  \item 12 parameters for the average mistag fractions $\mistag_i$ and the differences
        $\Delta\mistag_i$ between \Bz and \Bzb mistag fractions for each tagging
        category, 
  \item 1 parameter for the small difference between \Bz and \Bzb reconstruction 
        efficiency~\cite{Aubert:2002rg}, and
  \item 6 parameters for the small difference between \Bz and \Bzb tagging 
        efficiencies in each tagging category~\cite{Aubert:2002rg}.
\end{itemize}
The background parameters that are allowed to vary are 
\begin{itemize}
 \item 24 mistag fraction parameters: $\mistag_i$ and 
   $\Delta\mistag_i$ of each tagging category for background components with
   zero and non-zero lifetime, respectively. 
 \item 3 parameters for the \deltat resolution: $\delta_\mathrm{core}$, 
  $S_\mathrm{core}$ and $f_\mathrm{core}$.
 \item 4 parameters for the  \Bflav time dependence:
  2 parameters for the fraction ($f_\mathrm{prompt}$) of zero lifetime 
  component for the lepton-tagged and nonlepton-tagged events,
  $\tau_{\mathrm{bg}}$, and $\Delta m_{d,\mathrm{bg}}$. 
 \item 8 parameters for possible \CP violation in the background, 
       including the apparent \CP asymmetry of non-peaking events 
       in each tagging category,
 \item 1 parameter for possible direct \CP violation in
       the $\chicone\KS$ background coming from $\jpsi\Kstarz$, and 
 \item 3 parameters for possible direct \CP violation in the $\jpsi\KL$ mode, 
       coming from $\jpsi\KS$, $\jpsi\Kstarz$, and the remaining $\jpsi$ backgrounds.
\end{itemize}
The effective value of \abslambda\ for
 the non-\jpsi background is fixed from a fit to the \jpsi-candidate
sidebands in $\jpsi\KL$. We fix $\tau_{\Bz}=1.530\ps$ and 
$\Delta m_d = 0.507\ps^{-1}$~\cite{Yao:2006px}.
The determination of the mistag fractions and \deltat resolution
function parameters for the signal is dominated by the
\Bflav sample, which is about 10 times larger than the \CP sample.

\section{\boldmath LIKELIHOOD FIT VALIDATION\label{sec:fitvalidation}}

We perform three  tests to validate the fitting procedure.  
The first of these tests consists of
generating ensembles of simulated experiments from the probability density
function (PDF) and fitting each
simulated experiment. We determine that the fitted values of $S_f$ and $C_f$ 
parameters are unbiased, and that the fit returns reasonable estimates of the 
statistical uncertainties, by verifying the  distribution of the
pull $\cal P$ on a parameter ${\cal O}$, given by 
${\cal P} = ({\cal O}_{\mathrm{fit}} - {\cal O}_{\mathrm{gen}})/\sigma({\cal O}_{\mathrm{fit}})$,
is consistent with a Gaussian centered about zero with a width of one.  
The quantity ${\cal O}_{\mathrm{fit}}$ is the
fitted value, with a fitted error of $\sigma({\cal O}_{\mathrm{fit}})$, and
${\cal O}_{\mathrm{gen}}$ is the generated value.

The second test involves fitting simulated signal events that include 
the full \babar\ detector simulation.
For each decay mode, we divide the signal MC sample to many data-sized
samples, fit them one by one, and then examine the distribution of 
the fitted results.
We make sure that the $\cal P$ distributions for these signal-only
simulated experiments 
are consistent with a Gaussian distribution centered at zero with a
width of one.

The third test is to perform null tests on
control samples of neutral and charged \B events where $S_f$ and $C_f$
should be very small or zero.  
The parameters $S_f$ and $C_f$
are consistent with zero for the charged \B sample of
$\jpsi K^\pm$, $\psitwos K^\pm$, $\chicone K^\pm$, and $\jpsi
\Kstarpm$ final states. For the neutral \Bflav sample, we find that
the $S_f$ and $C_f$ parameters slightly deviate from zero at approximately
twice the statistical uncertainty (see Table~\ref{tab:result}). The
deviation of $S_f$ from zero is consistent with the directly measured 
\CP\ asymmetry $S\sim -2 r \sin(2\beta+\gamma)\cos(\delta)
\lesssim 0.04$~\cite{Long:2003wq} in $\Bz \to D^{(*)\pm} h^\mp$~\cite{Aubert:2006tw} 
due to interference from doubly-CKM-suppressed decays, where $\gamma = \arg [-(\vud\vub^*) /
  (\vcd\vcb^*)]$, $\delta$ is the strong phase difference between
CKM-favored and doubly-CKM-suppressed amplitudes, and $r \sim 0.02$ is
the ratio of the two amplitudes. Considering this expected \CP\
asymmetry in the \Bflav sample and systematic uncertainties (at $\sim
1\%$ level), we conclude that our analysis is free of pathological
behaviors.

\section{\boldmath RESULTS\label{sec:results}}
\begin{figure}[!htb]
\begin{center}
\includegraphics*[width=0.45\textwidth] {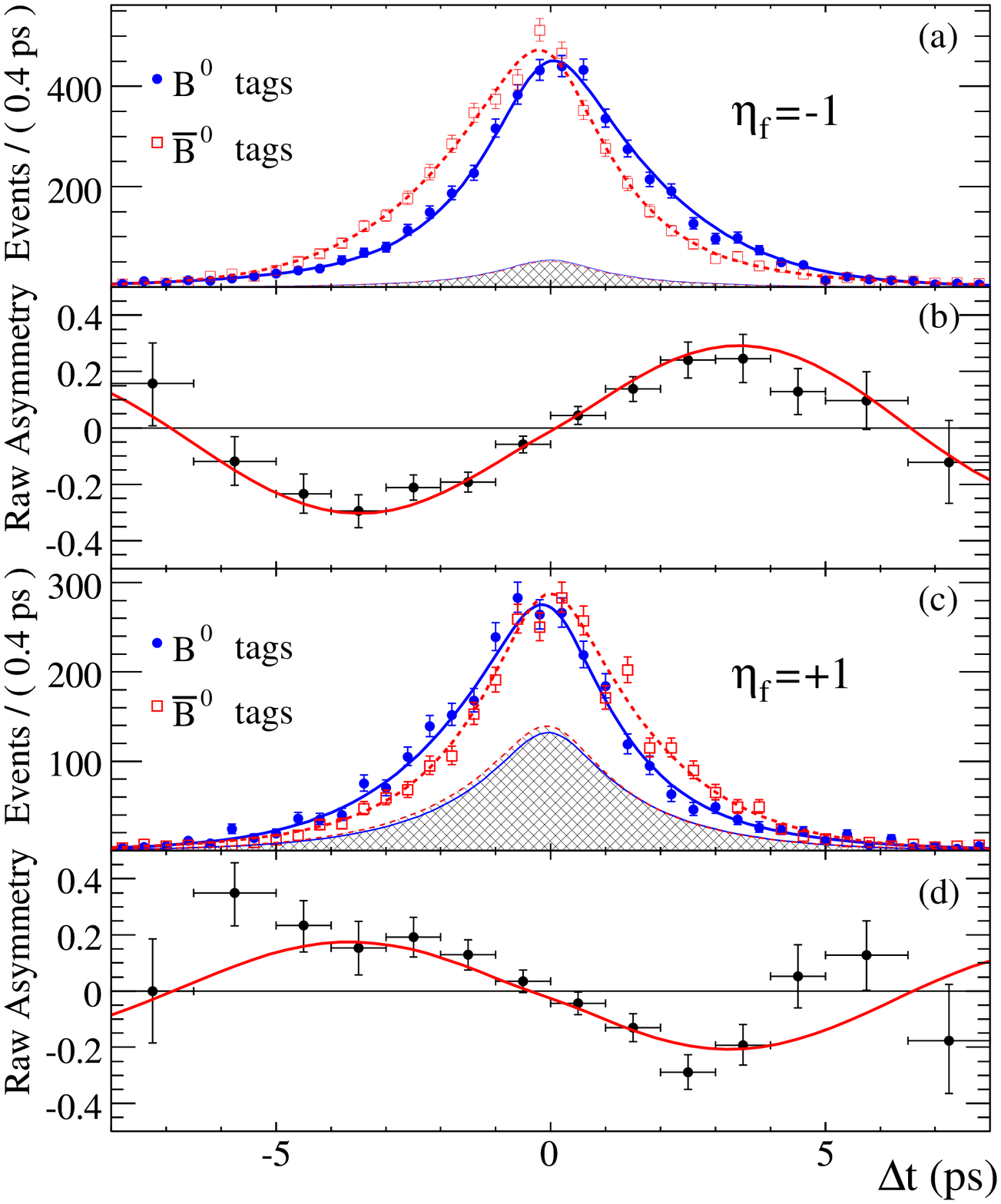}
\caption{
a) Number of $\eta_f=-1$ candidates ($\jpsi\KS$, $\psitwos\KS$,
$\chicone \KS$, and $\etac \KS$) in the signal region with a \Bz tag
($N_{\Bz }$) and with a \Bzb tag ($N_{\Bzb}$), and
b) the raw asymmetry, $(N_{\Bz}-N_{\Bzb})/(N_{\Bz}+N_{\Bzb})$, as functions
of \deltat;
 c) and d) are the corresponding distributions for the $\eta_f=+1$
mode $\jpsi\KL$.
The solid (dashed) curves in (a) and (c) represent the fit projections 
in \deltat for \Bz (\Bzb) tags. The shaded regions represent the 
estimated background contributions to (a) and (c). The curves in (b) and (d)
are the fit projections of the raw asymmetry between \Bz tagged and \Bzb tagged 
events.}
\label{fig:cpdeltat}
\end{center}
\end{figure}
\begin{figure}[!htb]
\begin{center}
\includegraphics*[width=0.46\textwidth] {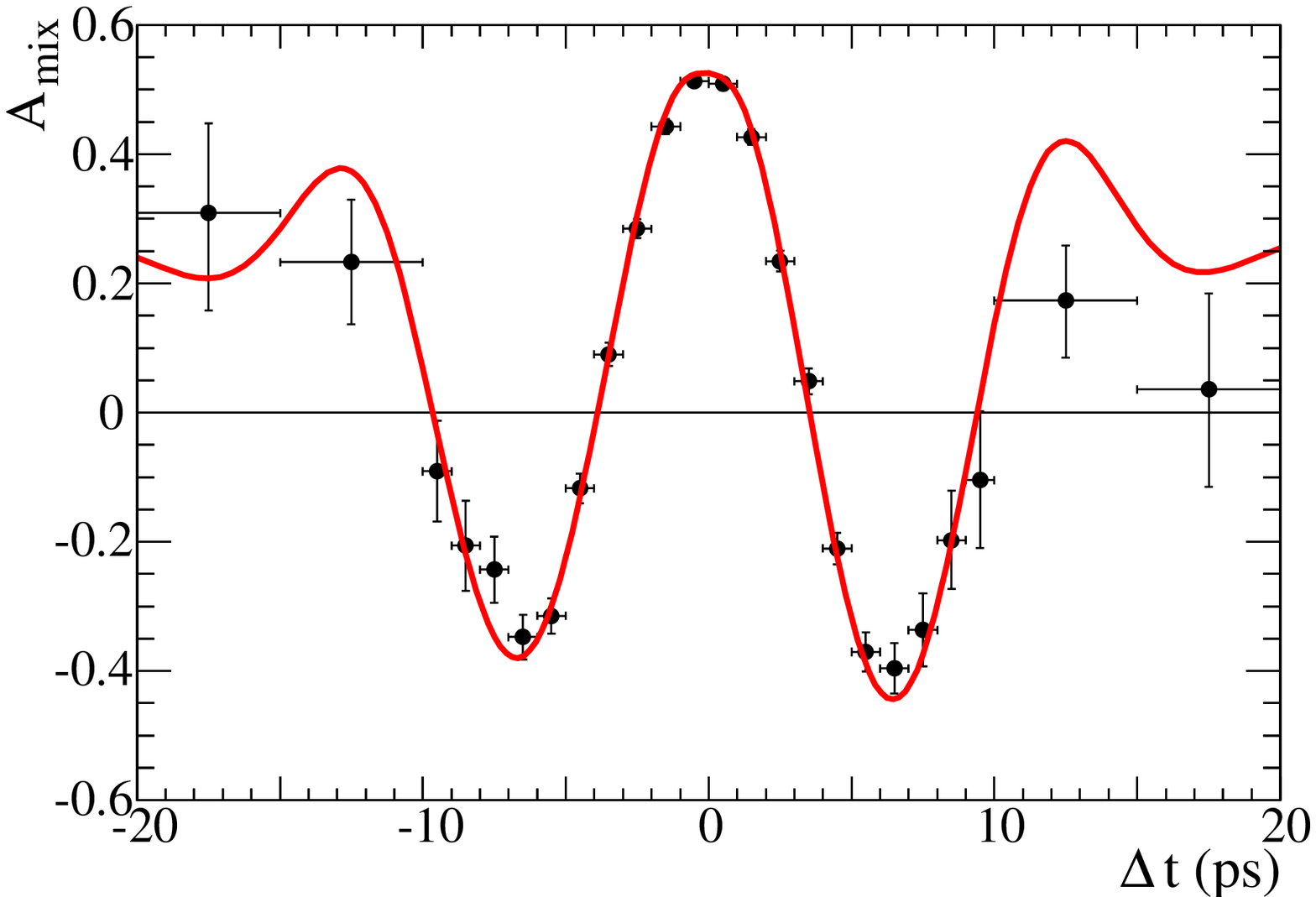}
\caption{Time-dependent asymmetry between unmixed and mixed events, 
$A_{\mbox{mix}}=(N_{\mbox{unmix}}- N_{\mbox{mix}})/(N_{\mbox{unmix}}+N_{\mbox{mix}})$,
as a function of \deltat\ for 
hadronic \B\ candidates (\Bflav) with $\mes>5.27\,\gevcc$.
The curve is the corresponding fit projection.
}
\label{fig:asym_bflav}
\end{center}
\end{figure}
\begin{figure*}[!htb]
\begin{center}
\includegraphics[width=1\textwidth]{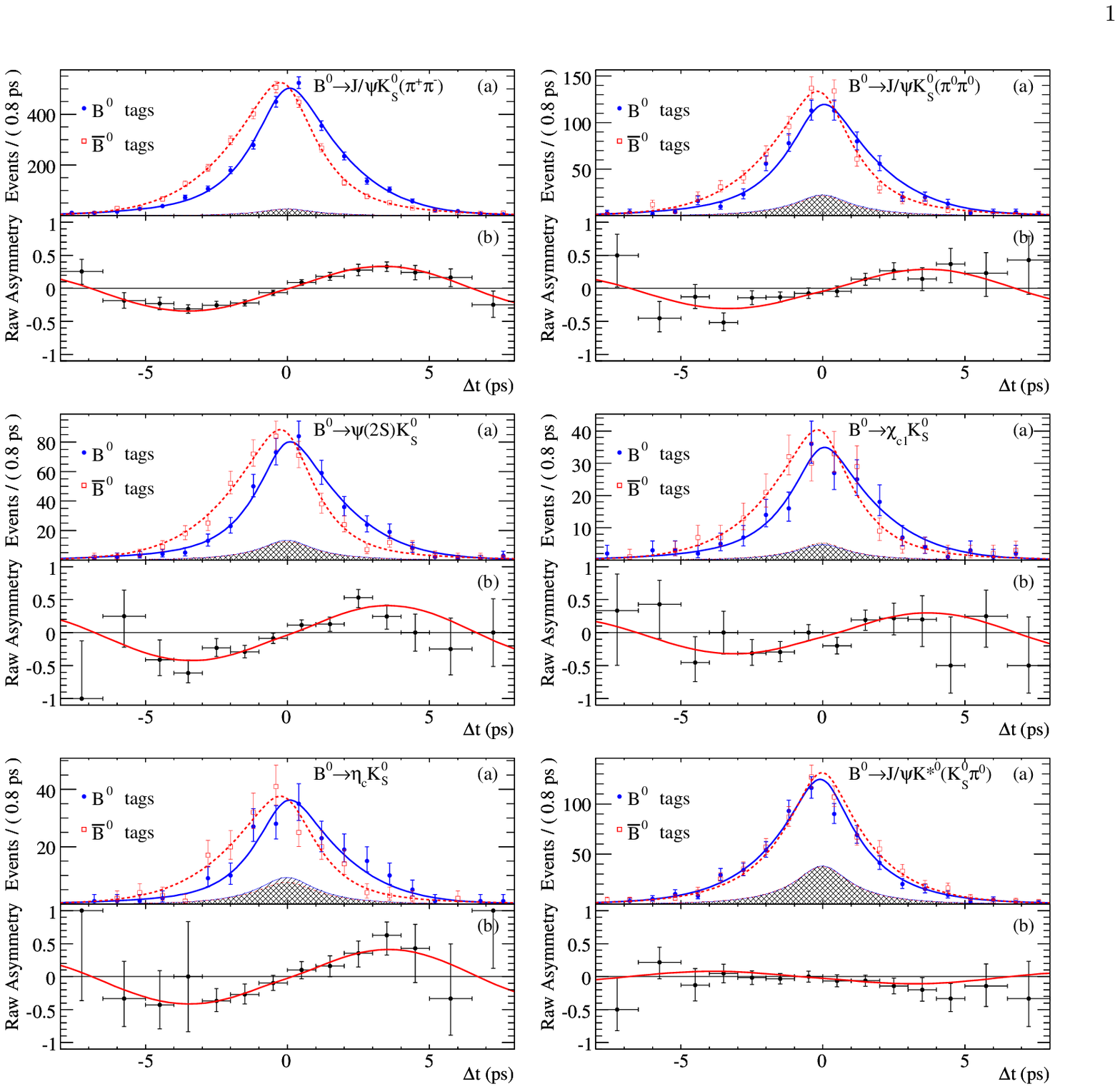}
\caption{
a) Number of $\Bz$ candidates in the signal region with a \Bz tag
($N_{\Bz }$) and with a \Bzb\ tag ($N_{\Bzb}$), and
b) the corresponding raw asymmetry, 
$(N_{\Bz}-N_{\Bzb})/(N_{\Bz}+N_{\Bzb})$, as functions
of \deltat\ for each \Bz\ decay modes.
The solid (dashed) curves represent the fit projections in \deltat for \Bz
(\Bzb) tags. The right-hatched (left-hatched) shaded regions represent the 
estimated background contributions in \Bzb (\Bz) tagged events.
}
\label{fig:cpdeltat_mode}
\end{center}
\end{figure*}

The fit to the \BCP and \Bflav samples yields $-\eta_f S_f = \fittedS$ and
$C_f = \fittedC$, where the errors are statistical only.
The correlation between these two parameters is $+\rhoSC\,\%$. 
We also performed the fit using \stwob and \abslambda as fitted parameters,
and found $\stwob=\fittedstwob$ and $\abslambda=\fittedmodl$.
The correlation between the fitted \stwob and \abslambda parameters
is $-0.14\,\%$.
Figure~\ref{fig:cpdeltat} shows the \deltat distributions and
asymmetries in yields between events with \Bz and \Bzb tags for the
$\eta_f=-1$ and $\eta_f = +1$ samples as a function of \deltat,
overlaid with the projection of the likelihood fit result.
Figure~\ref{fig:asym_bflav} shows the time-dependent asymmetry between 
unmixed and mixed events for hadronic \B\ candidates with $\mes>5.27\,\gevcc$.
We also perform a fit in which we allow different $S_f$ and $C_f$
values for each charmonium decay mode, a fit to the
$\jpsi\KS\,(\pipi+\ppz)$ mode, and a fit to the $\jpsi\Kz\,(\KS+\KL)$
sample. The results for some of these studies are 
shown in Figure~\ref{fig:cpdeltat_mode}.
We split the data sample by run period and by tagging
category. We perform the \CP measurements on control samples with
no expected \CP asymmetry. The results of these fits are summarized in
Table~\ref{tab:result}.

\begin{table*}[!htb]
\vskip-0.4truecm
\begin{center}
\caption{
Number of events $N_{\rm tag}$ and signal purity $P$ in the signal region
after tagging and vertexing requirements, and results of fitting for \CP
asymmetries in the \BCP sample and various subsamples.
Fit results for the \Bflav and $B^+$ control samples are also shown here.
Errors are statistical only.
}
\label{tab:result}
\begin{tabular*}{0.8\textwidth}{@{\extracolsep{\fill}}lrccc}\tbline\tbline
Sample  & $N_{tag}$ & \!$P(\%)$\! & $-\eta_f S_f$      & $C_f$ \\ \tbline
Full \CP sample & 15481 & 76  & \fittedS       & $\phantom{-}\fittedC$ \\ \tbline
\JpsiKsCh       & 5426 & 96   & \fittedSks     & $\phantom{-}\fittedCks$ \\
\JpsiKszz       & 1324 & 87   & \fittedSkszz   & $\phantom{-}\fittedCkszz$ \\
$\psitwos\KS$   & 861 &  87   & \fittedSpsitwoS& $\phantom{-}\fittedCpsitwoS$ \\
$\chicone\KS$   & 385 &  88   & \fittedSchic   & $\phantom{-}\fittedCchic$ \\
$\etac\KS $     & 381 &  79   & \fittedSetac   & $\phantom{-}\fittedCetac$ \\
$\jpsi\KL$      & 5813 & 56   & \fittedSkl     & $           \fittedCkl$ \\
$\jpsi\Kstarz$  & 1291 & 67   & \fittedSkst    & $\phantom{-}\fittedCkst$ \\ \tbline
$\jpsi\KS$      & 6750 & 95   & \fittedSjpsiks & $\phantom{-}\fittedCjpsiks$ \\ \tbline
$\jpsi\Kz$      & 12563 & 77  & \fittedSjpsikz & $\phantom{-}\fittedCjpsikz$ \\ \tbline
$\eta_f=-1$     & 8377 & 93   & $0.684\pm 0.032$ & $\phantom{-}0.037\pm 0.023$ \\ \tbline\hline
1999-2002 data  & 3079 & 78   & $0.732\pm 0.061$ & $\phantom{-}0.020\pm 0.045$  \\
2003-2004 data  & 4916 & 77   & $0.720\pm 0.050$ & $\phantom{-}0.045\pm 0.036$  \\
2005-2006 data  & 4721 & 76   & $0.632\pm 0.052$ & $\phantom{-}0.027\pm 0.037$  \\
2007 data       & 2765 & 75   & $0.663\pm 0.071$ & $-0.023\pm 0.049$ \\ \hline\hline
{\it Lepton}    & 1740 & 83   & $0.732\pm 0.052$ & $\phantom{-}0.074\pm 0.038$ \\
{\it Kaon I}    & 2187 & 78   & $0.615\pm 0.053$ & $-0.046\pm 0.039$ \\
{\it Kaon II}   & 3630 & 76   & $0.688\pm 0.056$ & $\phantom{-}0.068\pm 0.039$ \\
{\it KaonPion}  & 2882 & 74   & $0.741\pm 0.086$ & $\phantom{-}0.013\pm 0.061$ \\
{\it Pion}      & 3053 & 76   & $0.711\pm 0.132$ & $\phantom{-}0.016\pm 0.090$ \\
{\it Other}     & 1989 & 74   & $0.766\pm 0.347$ & $-0.176\pm 0.236$ \\ \tbline\tbline
\Bflav sample   & 166276 & 83 & $0.021\pm 0.009$  & $\phantom{-}0.012\pm 0.006$ \\
$B^+$ sample    & 36082  & 94 & $0.021\pm 0.016$  & $\phantom{-}0.013\pm 0.011$  \\\tbline\hline
\end{tabular*}
\end{center}
\end{table*}

The dominant systematic uncertainties on $S_f$ are summarized 
in Tables~\ref{tab:systematicssplitbymodeb} and~\ref{tab:systematicssplitbymode}.
The dilution due to flavor tagging can be different between \BCP  and
\Bflav events. We study this effect by comparing the results in
large samples of simulated \BCP and \Bflav events. The uncertainties due to
\deltat resolution functions for both signal and background components are
estimated by varying the fixed parameters and by using alternative models. We
also vary the peaking background fractions based on estimates derived from 
simulation, and vary the
\CP content of the background over a wide range to estimate the effect due to
our limited knowledge of background properties. The uncertainties in the 
$\jpsi\KL$ sample are studied by varying the 
compositions of the signal and background, 
by modifying the \deltae PDF based on studies performed 
with the $\jpsi\KS$ control sample, and
by varying the branching fractions of the background modes and their \CP
asymmetries. Other sources of uncertainty such as the values of
the physics parameters $\Delta m_d, \tau_B, \Delta \Gamma_d/\Gamma_d$, the
beam spot and detector alignment, and other fixed parameters, are studied by
varying them according to their world averages, the calibration, and the
statistical uncertainty, respectively. Despite the large amount of simulated
signal events that included the full \babar\ detector simulation, we
can only validate the possible fit bias to be no more than certain precision.
As a result, we assign a systematic uncertainty corresponding to any
deviations and the statistical uncertainties of the mean values of
the fitted $S_f$ and $C_f$ from the generated values as the possible 
fit bias (MC statistics).

The only sizable systematic uncertainties on $C_f$ are due to 
the \CP content of the peaking backgrounds and due to the
possible interference between the suppressed $\bar b\to \bar u c \bar
d$ amplitude with the favored $b\to c \bar u d$ amplitude for some tag-side
\B decays~\cite{Long:2003wq}.
The total systematic error on $S_f$ ($C_f$) is calculated by adding the
individual systematic uncertainties in quadrature and is found to be \systS\
(\systC).
The main sources of systematic uncertainty are listed in 
Tables~\ref{tab:systematicssplitbymodeb} and~\ref{tab:systematicssplitbymode}.

For the $\etac \KS$ mode, we found
$-\eta_f S_f = 0.925 \pm 0.160 \stat \pm 0.057 \syst$,  which has a significance 
of  $5.4\,\sigma$ standard deviations including systematic uncertainties. 
Our result is the first observation of \CP\ violation in this mode.

\begin{table*}[!htp]
\vskip-0.4truecm
\caption{Main systematic uncertainties on $S_f$ and $C_f$ for the
full \CP sample, and for the $\jpsi\Kz$, $\jpsi\KS$, and $\jpsi\KL$ samples. For each
source of systematic uncertainty, the first line gives the error on $S_f$
and the second line the error on $C_f$. The total systematic error
(last row) also includes smaller effects not explicitly mentioned in the
table.}
\def\KsCh{\ensuremath{K^{0\scriptstyle (+-)}_{\scriptscriptstyle S}}\xspace}
\def\Ksz{\ensuremath{K^{0\scriptstyle (00)}_{\scriptscriptstyle S}}\xspace}
\def\extParams{\ensuremath{\deltamd, \tau_B, \deltaGammad/\Gammad}\xspace}
\label{tab:systematicssplitbymodeb}
\begin{center}
\begin{tabular*}{0.8\textwidth}{@{\extracolsep{\fill}}lccccc}
\dbline
Source/sample & & Full
       & $\jpsi\Kz$
       & $\jpsi\KS$
       & $\jpsi\KL$ \\ \tbline
                                Beam spot & $S_f$ & 0.001 & 0.002 & 0.003 & 0.000 \\
                                         & $C_f$ & 0.001 & 0.001 & 0.002 & 0.000 \\\hline
                      Mistag differences & $S_f$ & 0.006 & 0.006 & 0.006 & 0.006 \\
                                         & $C_f$ & 0.002 & 0.002 & 0.002 & 0.002 \\\hline
                      \deltat resolution & $S_f$ & 0.007 & 0.007 & 0.007 & 0.007 \\
                                         & $C_f$ & 0.003 & 0.003 & 0.003 & 0.007 \\\hline
                   $\jpsi\KL$ background & $S_f$ & 0.006 & 0.006 & 0.000 & 0.027 \\
                                         & $C_f$ & 0.001 & 0.001 & 0.000 & 0.004 \\\hline
     Background fraction                 & $S_f$ & 0.005 & 0.004 & 0.004 & 0.004 \\
        and \CP content                  & $C_f$ & 0.003 & 0.002 & 0.001 & 0.011 \\\hline
                  \mes\ parameterization & $S_f$ & 0.002 & 0.002 & 0.003 & 0.001 \\
                                         & $C_f$ & 0.000 & 0.001 & 0.001 & 0.000 \\\hline
                              \extParams & $S_f$ & 0.003 & 0.003 & 0.004 & 0.004 \\
                                         & $C_f$ & 0.001 & 0.001 & 0.001 & 0.001 \\\hline
                   Tag-side interference & $S_f$ & 0.001 & 0.001 & 0.001 & 0.001 \\
                                         & $C_f$ & 0.014 & 0.014 & 0.014 & 0.014 \\\hline
                                Fit bias & $S_f$ & 0.002 & 0.004 & 0.004 & 0.006 \\
                         (MC statistics) & $C_f$ & 0.003 & 0.004 & 0.004 & 0.006 \\\hline
\hline                                     	
                                   Total & $S_f$ & 0.012 & 0.013 & 0.012 & 0.031 \\
                                         & $C_f$ & 0.016 & 0.018 & 0.016 & 0.027 \\\hline
\hline
\end{tabular*}
\end{center}
\end{table*}

\begin{table*}[!htp]
\vskip-0.4truecm
\caption{Main systematic uncertainties on $S_f$ and $C_f$ 
for the \JpsiKsCh, \JpsiKszz, $\psitwos\KS$, $\chicone\KS$,
$\etac\KS$, and $\jpsi\Kstarz (\Kstarz \to \KS\piz)$ decay
modes. For each
source of systematic uncertainty, the first line gives the error on $S_f$
and the second line the error on $C_f$. The total systematic error
(last row) also includes smaller effects not explicitly mentioned in the
table.}
\def\KsCh{\ensuremath{K^{0\scriptstyle (+-)}_{\scriptscriptstyle S}}\xspace}
\def\Ksz{\ensuremath{K^{0\scriptstyle (00)}_{\scriptscriptstyle S}}\xspace}
\def\extParams{\ensuremath{\deltamd, \tau_B, \deltaGammad/\Gammad}\xspace}
\label{tab:systematicssplitbymode}
\begin{center}
\begin{tabular}{@{\extracolsep{\fill}}lcccccccc}
\dbline
Source/sample & 
       & $\jpsi \KS {(\pipi)}$
       & $\jpsi \KS {(\ppz)}$
       & $\psi(2S) \KS$
       & $~~~\chicone \KS~~~$
       & $~~~\etac\KS~~~$
       & $~~~\jpsi\Kstarz~~~$ \\ \tbline
                                Beamspot &$S_f$ &  0.003 & 0.002 & 0.008 & 0.028 & 0.001 & 0.006 \\ 
                                         &$C_f$ &  0.002 & 0.003 & 0.009 & 0.012 & 0.000 & 0.000 \\ \hline 
                      Mistag differences & $S_f$ & 0.006 & 0.006 & 0.006 & 0.006 & 0.006 & 0.006 \\
                                         & $C_f$ & 0.002 & 0.002 & 0.002 & 0.002 & 0.002 & 0.002 \\ \hline
                      \deltat resolution &$S_f$ &  0.007 & 0.007 & 0.007 & 0.010 & 0.016 & 0.026 \\ 
                                         &$C_f$ &  0.003 & 0.004 & 0.007 & 0.004 & 0.004 & 0.006 \\ \hline 
                   $\jpsi\KL$ background &$S_f$ &  0.000 & 0.000 & 0.000 & 0.000 & 0.000 & 0.000 \\ 
                                         &$C_f$ &  0.000 & 0.000 & 0.000 & 0.000 & 0.000 & 0.000 \\ \hline 
     Background fraction                 &$S_f$ &  0.003 & 0.007 & 0.016 & 0.017 & 0.051 & 0.056 \\ 
                      and \CP content    &$C_f$ &  0.001 & 0.003 & 0.006 & 0.010 & 0.019 & 0.026 \\ \hline 
                  \mes\                  &$S_f$ &  0.002 & 0.009 & 0.024 & 0.006 & 0.002 & 0.037 \\ 
                   parameterization      &$C_f$ &  0.001 & 0.006 & 0.001 & 0.002 & 0.001 & 0.008 \\ \hline 
                              \extParams &$S_f$ &  0.003 & 0.007 & 0.016 & 0.003 & 0.016 & 0.014 \\ 
                                         &$C_f$ &  0.001 & 0.001 & 0.001 & 0.001 & 0.002 & 0.001 \\ \hline 
                   Tag-side interference &$S_f$ &  0.001 & 0.001 & 0.001 & 0.001 & 0.001 & 0.001 \\ 
                                         &$C_f$ &  0.014 & 0.014 & 0.014 & 0.014 & 0.014 & 0.014 \\ \hline 
                                Fit bias &$S_f$ &  0.005 & 0.004 & 0.008 & 0.007 & 0.007 & 0.027 \\ 
                         (MC statistics) &$C_f$ &  0.004 & 0.003 & 0.002 & 0.004 & 0.007 & 0.039 \\ \hline 
 \hline                                   	
                                   Total & $S_f$ & 0.012 & 0.017 & 0.036 & 0.040 & 0.057 & 0.087 \\
                                         & $C_f$ & 0.016 & 0.018 & 0.020 & 0.025 & 0.029 & 0.054 \\ \hline
 \hline
\end{tabular}
\end{center}
\end{table*}

\section{\boldmath CONCLUSIONS\label{sec:conclusions}}
\label{sec:conclusion}

We report improved measurements of the time-dependent \CP\ asymmetry
parameters. The results in this paper supercede those of our previous 
publication~\cite{Aubert:2007hm}. We
report our measurements in terms of $C_f$ and $S_f$. We find
\begin{eqnarray}
C_f &=& \measuredc,\nonumber\\
-\eta_f S_f &=& \measureds,\nonumber
\end{eqnarray}
providing an independent constraint on the position of the apex of
the Unitarity Triangle~\cite{ref:unitaritytriangle}.  Our measurements
agree with previous published results~\cite{Aubert:2007hm,Chen:2006nk}
and with the theoretical estimates of the magnitudes of CKM matrix
elements within the context of the SM~\cite{Ciuchini:1995cd}.  We also report measurements of
$C_f$ and $S_f$ for each decay mode in our \CP\ sample and
for the combined $\jpsi\Kz (\KS+\KL)$ mode. \CP violation in 
$\etac \KS$ mode is established at the level of $5.4\sigma$ standard deviation 
including systematic uncertainties. 

\section{ACKNOWLEDGMENTS}

We are grateful for the extraordinary contributions of our \pep2\ 
colleagues in achieving the excellent luminosity and machine conditions
that have made this work possible. The success of this project also 
relies critically on the expertise and dedication of the computing 
organizations that support \babar. The collaborating institutions 
wish to thank SLAC for its support and the kind hospitality extended 
to them. This work is supported by the US Department of Energy
and National Science Foundation, the Natural Sciences and Engineering 
Research Council (Canada), the Commissariat \`a l'Energie Atomique and
Institut National de Physique Nucl\'eaire et de Physique des Particules
(France), the Bundesministerium f\"ur Bildung und Forschung and
Deutsche Forschungsgemeinschaft (Germany), the Istituto Nazionale di 
Fisica Nucleare (Italy), the Foundation for Fundamental Research on 
Matter (The Netherlands), the Research Council of Norway, the
Ministry of Education and Science of the Russian Federation, 
Ministerio de Educaci\'on y Ciencia (Spain), and the
Science and Technology Facilities Council (United Kingdom).
Individuals have received support from  the Marie-Curie IEF 
program (European Union) and the A. P. Sloan Foundation.


\begin{thebibliography}{99}
\bibitem{ref:chargeconj}
Charge conjugation is implied throughout this paper.
\end{thebibliography}
\end{document}